\documentclass[prd,superscriptaddress,preprint]{revtex4}

\usepackage{mathrsfs,amsmath}
\usepackage{eurosym}
\usepackage{amssymb}
\usepackage{bbold}

\usepackage{latexsym,epsfig,epstopdf}
\usepackage{amssymb,amsmath,amsthm}

\usepackage{times}
\usepackage{color}

\usepackage{lmodern}
\usepackage[utf8]{inputenc}

\usepackage{graphicx}

\usepackage{color}

\usepackage{braket}

\usepackage[mathscr]{euscript} 

\def\be{\begin{equation}}
\def\ee{\end{equation}}
\def\bea{\begin{eqnarray}}
\def\eea{\end{eqnarray}}

\begin{document}

\title{Geometric criterion of topological phase transition for non-Hermitian systems}
\author{Annan Fan}
\affiliation{School of Physics, Sun Yat-Sen University, Guangzhou, 510275, China}
\author{Shi-Dong Liang }
\altaffiliation{Email: stslsd@mail.sysu.edu.cn}
\affiliation{School of Physics, Sun Yat-Sen University, Guangzhou, 510275, China}
\affiliation{State Key Laboratory of Optoelectronic Material and Technology, and\\
Guangdong Province Key Laboratory of Display Material and Technology,
Sun Yat-Sen University, Guangzhou, 510275, China}

\date{\today }

\begin{abstract}
We propose a geometric criterion of the topological phase transition for non-Hermitian systems.
We define the length of the boundary of the bulk band in the complex energy plane for non-Hermitian systems. For one-dimensional systems, we find that the topological phase transition occurs when the derivatives of the length with respect to parameters are discontinuous.
For two-dimensional systems, when the length is discontinuous, the topological phase transitions between the gapped and gapless phases occurs.
When the derivatives of the length with respect to parameters are discontinuous, the topological phase transition between the gapless and gapless phases occurs.
These nonanalytic behaviors of the length in the complex energy plane provide a signal to detect the topological phase transitions.
We demonstrate this geometric criterion by the one-dimensional non-Hermitian Su-Schieffer-Heeger model and the two-dimensional non-Hermitian Chern insulator model. This geometric criterion provides an efficient insight to the global topological invariant from a geometric local object in the complex energy plane for non-Hermitian systems.
\end{abstract}

\maketitle



\section{Introduction}
As theoretical models of dissipative or open systems, non-Hermitian quantum models have attracted growing attention in condensed matter physics due to their novel quantum characteristics and potential applications in quantum technology.\cite{Ramy,Jin,Gong,Kohei} In particular, it has been found that there exist richer topological equivalent classes in non-Hermitian systems than the conventional Hermitian systems based on symmetries and dimensions of quantum systems.\cite{Kohei,Kane,Chiu}
This yields a lot of fundamental issues, such as how to classify the topological phases, how to detect the topological phase transitions and what physical observables behind the robust of quantum states. It has been proved that the symmetry and gap structures of non-Hermitian systems are remained under a unitary flatterning  and Hermitianization transformation, \cite{Kohei} which promises a great potential to develop some quantum devices based on these robust quantum properties.

The topological phases can be characterized by the winding number and Chern number combined with the bulk-boundary correspondence for Hermitian systems.\cite{Kohei2,Ghatak} For non-Hermitian systems, the non-Hermiticity of quantum systems describes open or dissipative properties with gain or loss energy and particle fluctuation of system.\cite{Ramy,Gong,Chernyak} The non-Hermitian system exhibits unconventional characteristics, such as complex energy bands and nonorthogonal eigenstates,\cite{Dorje1,Berry} which lead to a lot of novel mathematical structures of quantum mechanics and rich physical phenomena, such as using the pseudo-Hermitian concept to reformulate the canonical domain for non-Hermitian systems. \cite{Ali,Hossein,Liang} The complex energy bands for non-Hermitian systems shows rich gap structures and complex exceptional points, such as the point and line gaps, which are associated with the topological phases.\cite{Kohei} One proposed the vorticity of the complex energy bands to characterize the topological phases.\cite{Ghatak,Fu} The bulk-boundary correspondence can be generalized to non-Hermitian systems by a similarity transformation even though the non-Hermiticity deforms the Bloch wave.\cite{Yin,Lieu,Rui,Tony,Daniel,Flore,Shunyu} The transport properties of non-Hermitian systems, such as quantum Hall conductance, can be generalized to quantum Hall admittance. \cite{Annan}

The phase boundaries between different topological phases in the parameter space are analyzed based on the exceptional points of the energy bands with the parameters varying for both Hermitian and non-Hermitian systems, which depends on the concrete model.\cite{Ghatak}
Recently, we found that the boundary of the bulk bands in the complex energy plane dominate topological invariants as a generalized bulk-boundary correspondence for non-Hermitian systems.\cite{Annan} From the geometric and topological points of views, the topological phase of quantum states is robust for the parameter deformation, which implies that the robust topological states are related to some geometric variables driving the topological phase transition. In particular, it is worth studying the geometric objects on the energy scale, such as the length of the boundary of the bulk bands in the complex energy plane, which could characterize topological invariants of quantum states for non-Hermitian systems.

In this paper, we propose a geometric criterion to detect the topological phase transition for non-Hermitian systems based on the analytic behaviors of the length of the bulk bands in the complex energy plane. In Sec. II, we first review briefly the basic physics and the description of topological invariants for non-Hermitian systems, such as the biorthogonal basis, complex energy bands, complex Berry phase, complex Chern number and vorticity for non-Hermitian systems. And then we
introduce the basic geometry of quantum states in the complex energy plane. We define a map from the tensor product space of the Brillouin zone and the parameter space to the complex energy plane and define the length of the bulk band in the complex energy plane.
We propose the geometric criterion based on the analytic behavior of the length in the complex energy plane to capture the topological phase transition.
In Sec. III, we demonstrate this geometric criterion by two typical non-Hermitian models, Su-Schrieffer-Heeger (SSH) and Chern insulator models.
We give the analytic expression of the length and its derivatives with respect to parameters for the non-Hermitian SSH model and demonstrate analytically and numerically the topological phase boundary based on this geometric criterion, which is consistent with those obtained by the conventional method.\cite{Kohei2,Ghatak,Fu}
We also demonstrate numerically the topological phase diagram of the non-Hermitian Chern insulator model based on this geometric criterion, which is consistent with those obtained by the conventional method.\cite{Kohei2,Ghatak,Fu} We discuss what physics behind the geometric criterion in Sec. IV. Finally we give the conclusions and outlooks in Sec. V.

\section{Geometric criterion of topological phase transition}

\subsection{Description of topological invariants for non-Hermitian systems}
The topological invariants of quantum states in Hermitian systems can be described by geometric and topological methods, such as Berry phase, winding number and Chern number. \cite{Ghatak}
These methods can be generalized to characterize topological invariants for non-Hermitian systems as long as the orthogonal basis is generalized to the biorthogonal basis and the eigenvalues are extended to the complex domain for non-Hermitian systems.\cite{Ghatak}

Let us first review briefly the basic physics and the topological phase of the non-Hermitian systems for comparing with that by our geometric method. Suppose that
a non-Hermitian system described by a non-Hermitian Hamiltonian, $H^{\dag}\neq H$ and the Hilbert space is separable, the Bloch Hamiltonian $H(\mathbf{k},\lambda)$ works in the Brillouin zone (BZ) with a parameter space, where $\mathbf{k}\in BZ^{d}$ is the d-dimensional Brillouin zone and $\lambda\in \mathcal{M}^{p}_{\lambda}$ is a p-dimensional parameter space. The eigen equations of the non-Hermitian Hamiltonian and its Hermitian adjoint are given by  \cite{Ghatak,Dorje1}
\begin{eqnarray}\label{HH}
H(\mathbf{k},\lambda)|\psi_{n}^{R}(\mathbf{k},\lambda)\rangle=E_{n}(\mathbf{k},\lambda)|\psi_{n}^{R}(\mathbf{k},\lambda)\rangle \\
H^{\dagger}(\mathbf{k},\lambda)|\varphi_{n}^{L}(\mathbf{k},\lambda)\rangle=E^{*}_{n}(k,\lambda)|\varphi_{n}^{L}(\mathbf{k},\lambda)\rangle.
\end{eqnarray}
The eigen vectors of the Hamiltonian and its adjoint are constructed to the biorthogonal basis of the separable Hilbert space,
\begin{equation}\label{Orbs}
\langle\varphi_{m}^{L}(\mathbf{k},\lambda)|\psi_{n}^{R}(\mathbf{k},\lambda)\rangle=\delta_{mn},
\end{equation}
and the complete relation is given by
\begin{equation}\label{CPR1}
\sum_{n}|\psi_{n}^{R}(\mathbf{k},\lambda)\left\rangle\right\langle\varphi_{n}^{L}(\mathbf{k},\lambda)|=I,
\end{equation}
where $I$ is the identity matrix. The complex Berry phase is defined as \cite{Ghatak}
\begin{equation}\label{BP1}
\gamma_{n}=\frac{1}{2\pi}\oint_\mathcal{C}\mathcal{A}_{n}\cdot d\mathbf{k},
\end{equation}
where $\mathcal{A}_{n}=i\langle\varphi^{L}_{n}(\mathbf{k},\lambda)|\nabla_{\mathbf{k}}\psi_{n}^{R}(\mathbf{k},\lambda)\rangle$ is the Berry connection. In general, the Berry phase in (\ref{BP1}) is complex for non-Hermitian systems. As the contour $\mathcal{C}$ encloses an exceptional point, $\gamma_{n}$ leads to a finite winding number, in which the real part gives the usual geometric phase acquired in each cycle, and is topological invariant for the specific cycle of the parameters as long as number of the exceptional points inside the contour remains the same. The corresponding imaginary part gives the decay part of the probability, and is not necessarily topological invariant.\cite{Ghatak,Fu}
The complex Berry phase in  (\ref{BP1}) can be converted into a surface integral and is defined as Chern number\cite{Ghatak,Fu}
\begin{equation}\label{CN1}
C_{n}=\frac{1}{2\pi}\int_\textrm{BZ}\Omega_{n} \cdot d\mathbf{S},
\end{equation}
where $\mathbf{\Omega}_{n}=\nabla\times \mathcal{A}_{n}$ is the Berry curvature. $d\mathbf{S}$ is the area section within the Berry flux.
In general the Chern number defined in (\ref{CN1}) are complex.\cite{Ghatak,Fu} The exceptional points of the complex energy bands give complex poles of the Berry curvature and represent effective complex monopoles in the BZ.

The topological phases are characterized by the topological indexes, such as winding number and Chern number. They are topological invariant under the parameter deformation. \cite {Chiu}
These topological indexes are equivalent to the topological edge states based on the bulk-boundary correspondence
even though for non-Hermitian systems,\cite {Chiu} in which the non-Hermiticity breaks the translation symmetry and deforms the Bloch wave. The bulk-boundary correspondence can be recovered by a similarity transformation.\cite{Yin,Lieu,Rui} Note that the energy bands are in general complex for non-Hermitian systems, the vorticity to characterize the topological phase is introduced by,\cite{Ghatak,Fu}
\begin{equation}\label{V1}
\nu_{n}=\frac{1}{2\pi}\oint\nabla_\mathbf{k} \phi_{n}\cdot d\mathbf{k}
\end{equation}
where $\phi_{n}=\arctan \frac{E_{nI}}{E_{nR}}$, where $E_{nR(L)}$ are the real and imaginary parts of the complex energy band.\cite{Ghatak, Fu}
The vorticity counts the winding number of the exceptional points in the complex energy plane, which inspires us to seek for some geometric variables related to the vorticity in the complex energy plane to characterize the topological phase for non-Hermitian systems.

In general, the topological phase based on the symmetry and dimension of systems is protected by the energy band gap. The Chern number characterizes the gapped-mode topological phase. When the energy band gap closes the Chern number becomes ill-defined. Thus, as the parameters vary, the closure of the energy band gap induces the topological phase transition. For the gapless modes, the vorticity plays a winding number role in the complex energy plane to characterize the topological phase.

\subsection{Complex energy plane and its geometry for non-Hermitian systems}
\begin{figure}[htbp]
	\centering
	\includegraphics[width=1\linewidth]{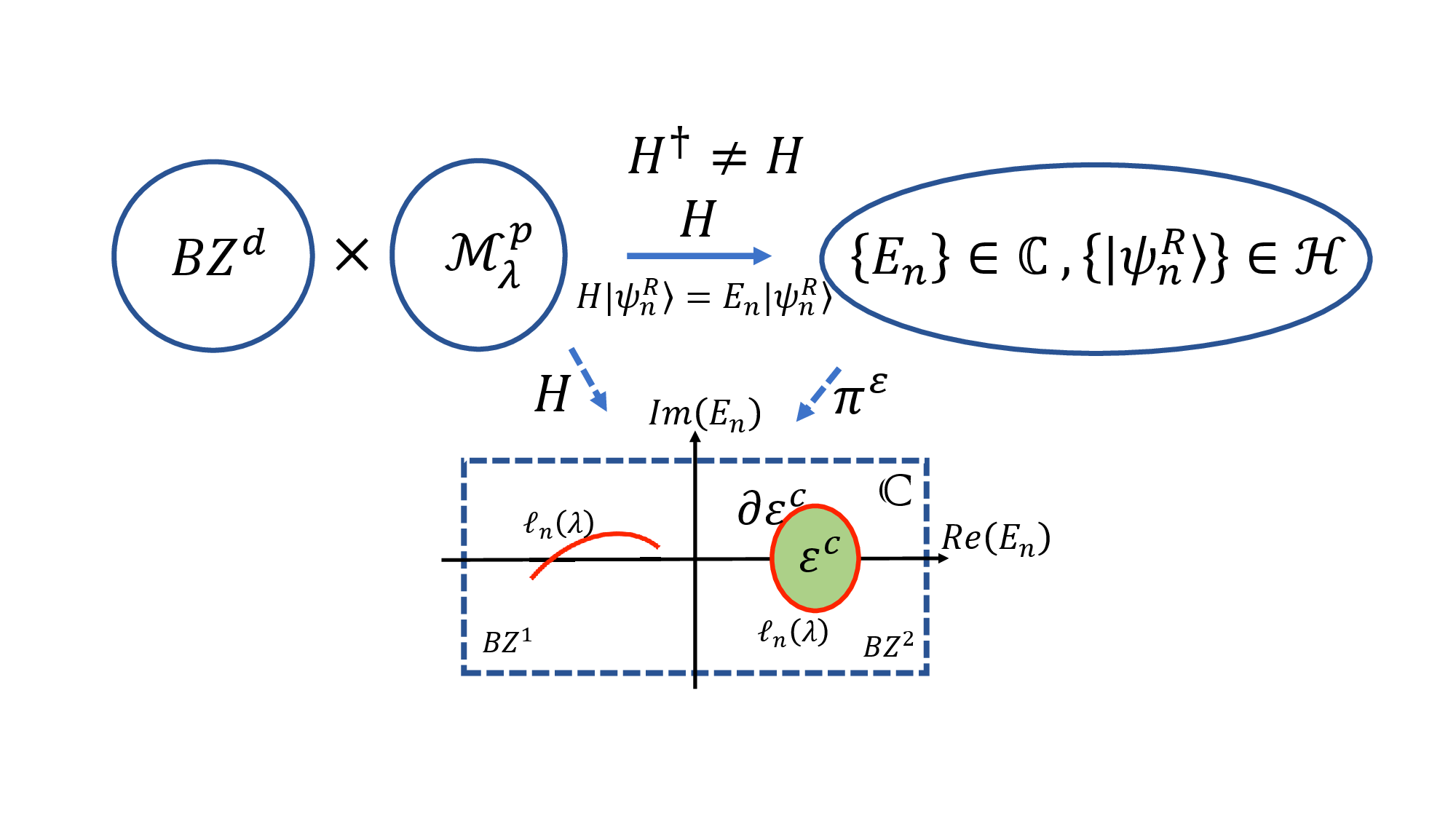}
	\caption{Online color: The schematic representation of the non-Hermitian Hamiltonian mapping $\pi^\varepsilon$ the complex energy band to the complex energy plane $\mathbb{C}$, in which $\varepsilon^c$ is the bulk band of the complex energy band. The red curve on the right side of the complex energy plane is the edge states of the bulk band $\partial \varepsilon^c$ for the 2D BZ. The red curve on the left side of the complex energy plane is the bulk band of the complex energy band in the complex energy plane for the 1D BZ.}
	\label{fig1}
\end{figure}

Note that
the energy bands $\left\{E_{n}(\mathbf{k},\lambda)\right\}$ are in general complex for non-Hermitian systems, we define a map $\pi^\varepsilon$ that maps the complex energy band to the complex energy plane denoted by $\pi^\varepsilon : E_{n}(\mathbf{k},\lambda)\rightarrow\left(E_{n}^{R}, E_{n}^{I}\right)\in\mathcal{E}^{c}(\mathbf{k},\lambda)$, where $\left(E_{n}^{R}, E_{n}^{I}\right)$ are the real and imaginary parts of the complex energy band respectively. The schematic illustration is given in Fig.\ref{fig1}.

For given a specific parameter $\lambda\in \mathcal{M}_{\lambda}^{p}$, the complex energy band in the BZ mapped to the complex energy plane form a bulk band in the complex energy plane. For the 1D BZ the bulk band in the complex energy plane is a curve (red curve on the left side of the complex energy plane in Fig. (\ref{fig1})) whereas for the 2D BZ the bulk band in the complex energy plane are some 2D regions in general (green region on the right side of the complex energy plane in Fig. (\ref{fig1})). Thus.
we define the length of the curve in the complex energy plane for the complex energy band,
\begin{equation}\label{LL0}
\ell_{n}(\lambda):=\left\{
\begin{array}{c}
\int_{\mathcal{C}\in \mathcal{E}^{c}} d\ell_{\varepsilon_{n}^c}(\lambda),  \quad \textrm{for} \quad BZ^1 \\
\int_{\partial \mathcal{E}^{c}}d\ell_{\varepsilon_{n}^c}(\lambda), \qquad \textrm{for} \quad BZ^{2},
\end{array}
\right.
\end{equation}
where $\mathcal{C}$ is the curve in the complex energy plane for the cases of the 1D BZ, and $\partial \mathcal{E}^{c}$ is the boundary of the bulk band in the complex energy plane for the 2D BZ. As the parameters vary, the curve moves in the complex energy plane for the 1D BZ.  The shape of the bulk band also changes in the complex energy plane for the 2D BZ leading to the boundary of the bulk band varying. The analytic behavior of the lengths in the complex energy plane provides the geometric criterion of the topological phase transition.

It should be remarked that the definition of the length in (\ref{LL0}) is well-defined when the wave vector in the BZ is continuous. In practice, we calculate the length by the numerical method (ball pivoting algorithm see the end of section III), in which the approximation of the length depends on the discrete precision of the BZ and the input of the radius of the ball pivoting algorithm. However, the approximation of length affects only the precision of the length of the curve, but does not affect the analytic behaviors of the length when the parameters cross the boundary of the phase transition. In other words, the numerical method and its approximation does not change the results because the topological invariant is robust and the numerical method is efficient.

\subsection{Geometric criterion of topological phase transitions}
Suppose that the topological phase diagram of the non-Hermitian systems is represented by a union of subspaces in the parameter space, $\Lambda=\bigcup_{\alpha}\Lambda_{\alpha}$, where $\lambda_{\alpha}\in \Lambda_\alpha$ denotes the topological phase $\alpha$. In general, the length of the curve defined in (\ref{LL0}) varies with variation of the parameter. We investigate the analytic behaviors of the length for the non-Hermitian SSH and Chern insulator models. We find that the topological phase transition can be detected by the analytic behaviors of the length in the complex energy plane. Thus, we propose a geometric criterion to detect the topological phase transition for non-Hermitian systems.

\textbf{Geometric criterion of topological phase transition:} For a non-Hermitian Dirac model, $H(\mathbf{k},\lambda)=\mathbf{h}(\mathbf{k},\lambda)\cdot\sigma$, where $\mathbf{h}(\mathbf{k},\lambda)\in \mathbb{C}$, the topological phase transition occurs at the nonanalytic behaviors of the length defined by (\ref{LL0}) in the parameter space.
\begin{itemize}
  \item For the cases of 1D $BZ^1$, the topological phase transition occurs at the discontinuous derivations of the length with respect to parameters.
  \item For the cases of 2D $BZ^2$, there are two types of the boundaries in the phase diagram of the topological phase transitions.\\
  (a) The topological phase transitions between the gapless and gapless phases occur when the derivations of the length with respect to parameters is discontinuous.\\
  (b) The topological phase transition between the gapped and gapless phases occurs when the length is discontinuous with variation of parameters.
\end{itemize}

The analytic behaviors of the length of the curve in the complex energy plane provide a signal for the topological phase transition. We will demonstrate this criterion in the following sections by two typical non-Hermitian models.

\section{Topological phase diagrams based on the geometric criterion}
Let us consider two typical non-Hermitian models, one is the 1D non-Hermitinan SSH model and the other is the 2D non-Hermitian Chern insulator model in this section. We  compare the phase diagrams of these two models by the conventional topological indexes, winding number, Chern number and vorticity with those by the geometric criterion. We explain what physics behind this geometric criterion.

\subsection{Non-Hermitian Su-Schrieffer-Heeger model}
As a typical 1D non-Hermitian system, we review the phase diagram of the non-Hermitian SSH model, which has been studied extensively.\cite{Yin,Lieu,Rui} Let us recall the  basic results on the topological phase diagram. The Hamiltonian of the non-Hermitian SSH model is given by \cite{Ghatak}
\begin{equation}\label{CS1}
H(k)=\mathbf{h}(k)\cdot \mathbf{\sigma}
\end{equation}
where
\begin{eqnarray}\label{SSH}
h_{x}(k) &=& t-\delta+t'\cos k\\
h_{y}(k) &=& t+\delta+it'\sin k \\
h_{z}(k) &=& 0,
\end{eqnarray}
where $t,t',\delta\in \mathbb{R}$. $t$ and $t'$ denote the intracell and intercell hopings of electrons respectively. $\delta$ describes the Bloch electron energy gain and loss. The wave vector is within the BZ, $-\pi\leq k\leq \pi$. The eigen energies are obtained
\begin{equation}\label{17E}
E_{\pm}(k)=\pm\sqrt{t'^2+t^{2}-\delta^{2}+2t't\cos{k}-2it'\delta\sin{k}},
\end{equation}
which can be rewritten to $E_{\pm}=\pm (E_{R}+i\sigma E_{I})$ with the real and imaginary parts of the complex energy bands,
\begin{eqnarray}\label{RIE1}
E_R &=& \sqrt{\frac{\epsilon +\sqrt{\epsilon^2+\omega^2}}{2}} \\
E_I &=& \sqrt{\frac{-\epsilon +\sqrt{\epsilon^2+\omega^2}}{2}} ,
\end{eqnarray}
where
\begin{eqnarray}\label{RIE2}
\epsilon &=& t'^2+t^{2}-\delta^{2}+2t't\cos{k} \\
\omega &=& 2t'\delta\sin{k},
\end{eqnarray}
and $\sigma=1$ for $\omega\geq 0$ and $\sigma=-1$ for $\omega< 0$.

The topological phase transition occurs at the closure of the energy gap. Namely the phase boundary is determined by $E_{\pm}=0$ with variation of parameters. $E_I=0$ gives $k=0,\pi$. Thus, the boundaries of the topological phase transitions are $t=\pm\delta+t'$ and $t=\pm\delta-t'$ in the parameter space.\cite{Yin}
The topological phases can be characterized by the winding number,\cite{Yin}
\begin{equation}\label{WN1}
w=\frac{1}{2\pi}\oint\partial_k \phi dk
\end{equation}
where $\tan\phi=\frac{h_y}{h_x}$. It has been found that the winding numbers are $0,\frac{1}{2}$ and $1$ for different topological phases in the parameter space.\cite{Yin}
Actually, the winding number in (\ref{WN1}) is an analog with the vorticity in (\ref{V1}) for 1D systems.

Let us reexamine the topological phase diagram based on the geometric criterion, note that the length defined in (\ref{LL0}) for the 1D BZ can be given by
\begin{equation}\label{LL2}
\ell=\int_{-\pi}^{\pi}\sqrt{\left(\frac{\partial E_R}{\partial k}\right)^{2}+\left(\frac{\partial E_I}{\partial k}\right)^{2}}dk,
\end{equation}
Note that
\begin{eqnarray}\label{ERI3}
E_{R}^{2}-E_{I}^{2} &=& t'^2+t^{2}-\delta^{2}+2t't\cos{k}\\
E_{R}E_{I} &=& 2t'\delta\sin{k},
\end{eqnarray}
we have
\begin{eqnarray}
E_{R}\frac{\partial E_R}{\partial k}-E_{I}\frac{\partial E_I}{\partial k} &= -tt'\sin k, \label{DERI1a}\\
E_{I}\frac{\partial E_R}{\partial k}+E_{R}\frac{\partial E_I}{\partial k} &= \delta t'\cos k. \label{DERI1b}
\end{eqnarray}
We rewrite (\ref{DERI1a}) and (\ref{DERI1b})to the matrix form
\begin{equation}\label{DERIx2}
\left(
\begin{array}{cc}
E_{R} & -E_{I} \\
E_{I} & E_{R}
\end{array}
\right)
\left(
\begin{array}{c}
\frac{\partial E_R}{\partial k} \\
\frac{\partial E_I}{\partial k}
\end{array}
\right)=\left(
\begin{array}{c}
-tt'\sin k \\
\delta t'\cos k
\end{array}
\right),
\end{equation}
Thus, for the gapped regions, $E=\sqrt{E_{R}^{2}+E_{I}^{2}}\neq 0$, the matrix in (\ref{DERIx2}) is reversible.
The derivatives of the real and imaginary energy bands  with respect to $k$ can be expressed as
\begin{equation}\label{DERIx5}
\left(
\begin{array}{c}
\frac{\partial E_R}{\partial k} \\
\frac{\partial E_I}{\partial k}
\end{array}
\right)=\frac{1}{E^{2}}
\left(
\begin{array}{c}
-tt'E_{R}\sin k+t'\delta\cos k \\
tt'E_{I}\sin k+t'\delta\cos k
\end{array}
\right).
\end{equation}
Consequently by substituting the derivatives in (\ref{DERIx5}) into (\ref{LL2}), the length of the curve in the complex energy plane can be expressed as
\begin{equation}\label{LL3}
\ell=t'\int_{-\pi}^{\pi}\frac{\sqrt{t^2 \sin^{2}k+\delta^2 \cos^{2}k}}{E}dk.
\end{equation}

Similarly, we obtain
\begin{eqnarray}\label{DERItd1}
\left(
\begin{array}{c}
\frac{\partial E_R}{\partial t} \\
\frac{\partial E_I}{\partial t}
\end{array}
\right)&=& \frac{1}{E^{2}}
\left(
\begin{array}{c}
2E_{R}(t+t'\cos k )\\
-2E_{I}(t+t'\cos k)
\end{array}
\right), \\
\left(
\begin{array}{c}
\frac{\partial E_R}{\partial \delta} \\
\frac{\partial E_I}{\partial \delta}
\end{array}
\right)&=&\frac{1}{E^{2}}
\left(
\begin{array}{c}
-2\delta E_{R}+t'E_{I}\sin k \\
2\delta E_{I}+t'E_{R}\sin k
\end{array}
\right).
\end{eqnarray}

The derivatives of the length with respect to $t$ and $\delta$ are obtained
\begin{eqnarray}\label{LL4}
\frac{\partial \ell}{\partial t} &=& 2t'\int_{-\pi}^{\pi}\frac{(t+t'\cos k)(E^{2}_{R}-E^{2}_{I})}{E^{7/2}}\sqrt{t^2 \sin^{2}k+\delta^2 \cos^{2}k}dk \nonumber \\
&+& t't\int_{-\pi}^{\pi}\frac{\sin^{2}k}{E\sqrt{t^2 \sin^{2}k+\delta^2 \cos^{2}k}}dk, \\
\frac{\partial \ell}{\partial \delta} &=& t'\int_{-\pi}^{\pi}\frac{2\delta(E^{2}_{R}-E^{2}_{I})+t'E_{R}E_{I}\sin k)}{E^{2}}\sqrt{t^2 \sin^{2}k+\delta^2 \cos^{2}k}dk
\nonumber \\
&+& t'\delta\int_{-\pi}^{\pi}\frac{\cos^{2}k}{E\sqrt{t^2 \sin^{2}k+\delta^2 \cos^{2}k}}dk.
\end{eqnarray}
where
\begin{equation}\label{EM1}
E=\sqrt[4]{(t'^2+t^{2}-\delta^{2}+2t't\cos{k})^{2}+(2t'\delta\sin{k})^{2}}.
\end{equation}
At the exceptional points, $|E|\rightarrow 0$, namely, the length and its derivatives respect with $t$ and $\delta$ are divergent in the boundary of the topological phase transition. In fact, we also demonstrate the length and its derivative with respect to parameters are divergent in the boundary of the topological phase transition. Thus, we prove the geometric criterion.

To seek for the relationship between the geometric criterion based on the length defined in (\ref{LL2}) and the winding number defined in (\ref{WN1}), we make the transform of the length to its polar coordinate representation. Note that the energy band can be rewritten to its polar coordinate representation $E_{\pm}=|E|e^{i\phi_{\pm}}$, where
\begin{equation}\label{Phi1}
\tan\phi_{\pm}=\frac{E_{I}}{E_{R}}=\pm\sqrt{\frac{-1+\sqrt{1+(2t'\delta\sin{k})^{2}/(t'^2+t^{2}-\delta^{2}+2t't\cos{k})^{2}}}
{1+\sqrt{1+(2t'\delta\sin{k})^{2}/(t'^2+t^{2}-\delta^{2}+2t't\cos{k})^{2}}}}.
\end{equation}
The length in the polar coordinate system can be rewritten as
\begin{equation}\label{LL5}
\ell=\int_{-\pi}^{\pi}\sqrt{\left(\frac{\partial |E|}{\partial k}\right)^{2}+|E|^{2}\left(\frac{\partial \phi}{\partial k}\right)^{2}}dk,
\end{equation}
where we neglect the subindex $\pm$ in $\phi$ without loss of generality. Suppose that $\frac{\partial |E|}{\partial k} < |E|\frac{\partial \phi}{\partial k}$, we have
\begin{equation}\label{LL6}
\sqrt{\left(\frac{\partial |E|}{\partial k}\right)^{2}+|E|^{2}\left(\frac{\partial \phi}{\partial k}\right)^{2}}\approx
|E|\frac{\partial \phi}{\partial k}\left[1+\frac{\left(\frac{\partial |E|}{\partial \phi}\right)^{2}}{2|E|^{2}}  \right].
\end{equation}
Consequently, we obtain the length approximately
\begin{eqnarray}\label{LL7}
\ell &=& \int_{-\pi}^{\pi}\frac{\partial \phi}{\partial k}
\left(|E|+\frac{1}{2|E|} \left(\frac{\partial |E|}{\partial \phi}\right)^{2} \right) dk \nonumber \\
&\approx & 2\pi w \left(|E|+\frac{1}{2|E|}\left(\frac{\partial |E|}{\partial \phi}\right)^{2} \right),
\end{eqnarray}
where we have assumed that the factor $|E|+\frac{1}{2|E|}\left(\frac{\partial |E|}{\partial \phi}\right)^{2}$ is weakly dependent on $k$. It can be seen that the first term of the length in (\ref{LL7}) is proportional to the winding number in (\ref{WN1}) in the energy dimension.

In order to demonstrate further the length of the curve in the complex energy plane as a geometric indicator to detect the topological phase transition, we plot the phase diagram of the non-Hermitian SSH model in the parameter space numerically based on the geometric criterion in Sec. II, in which we set $t'=1$ for convenience and comparing with the previous results without loss of generality.\cite{Yin}

\begin{figure}[htbp]
	\centering
	\includegraphics[width=1\linewidth]{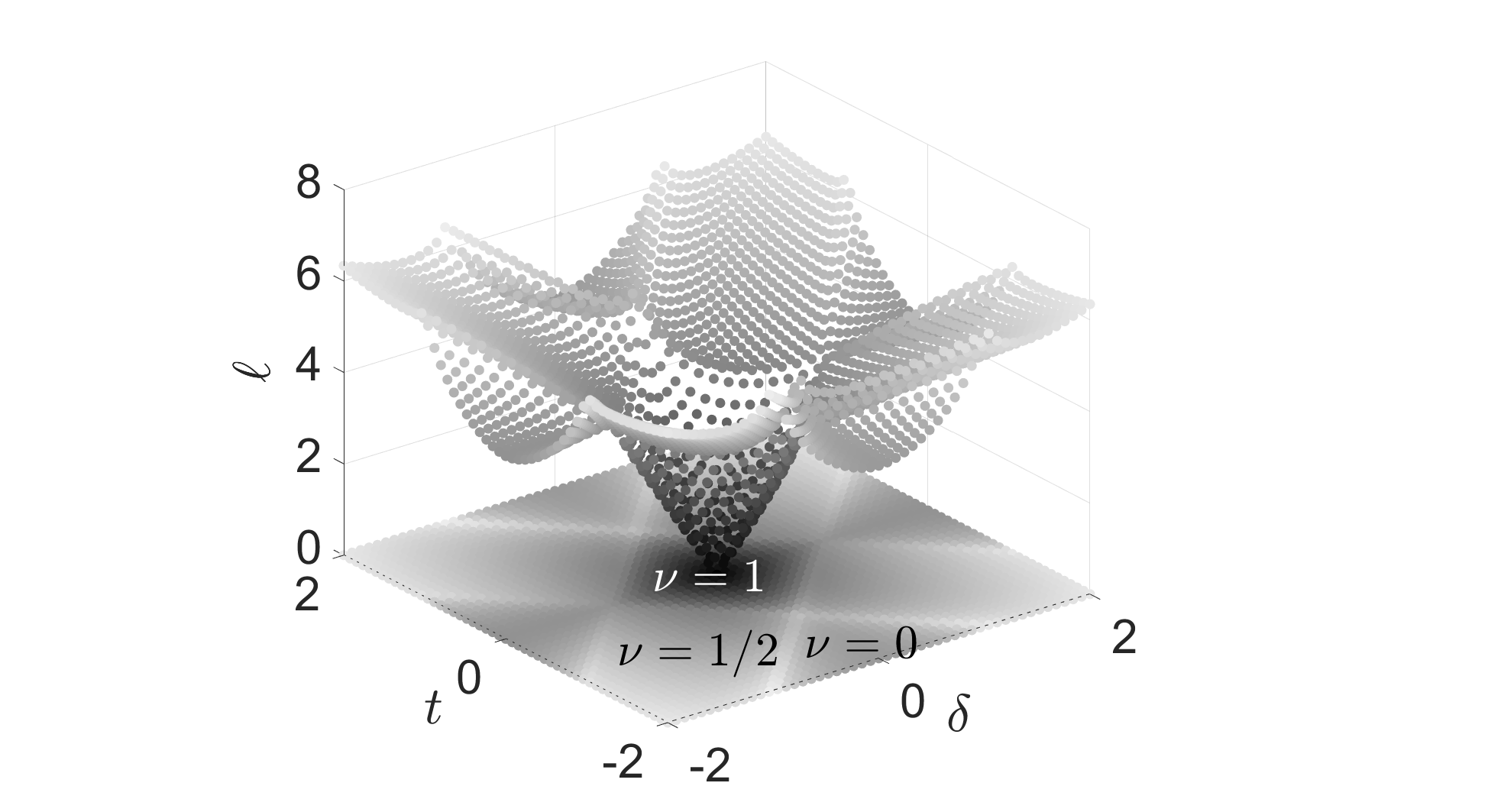}
	\caption{Online color: The length in the complex energy plane in the parameter space $t-\delta$ for given $t'=1$. The projection in the parameter space is exactly same the topological phase diagram determined by the exceptional points label by the winding number.}
	\label{fig2}
\end{figure}

The lengths of the curves based on (\ref{LL3}) in the parameter space are plotted in Fig.(\ref{fig2}) for the lower energy band. It can be seen that
the projection of the length in the parameter space is exactly same to the phase diagram determined by the exceptional point, $t=\pm \delta+t'$ and $t=\pm \delta-t'$.\cite{Yin} The boundary of the projection of the length on the parameter space stems from the derivations of the length in the complex energy plane with respect to the parameters $t$ and $\delta$ are discontinuous.

\begin{figure}[htbp]
	\centering
	\includegraphics[width=1\linewidth]{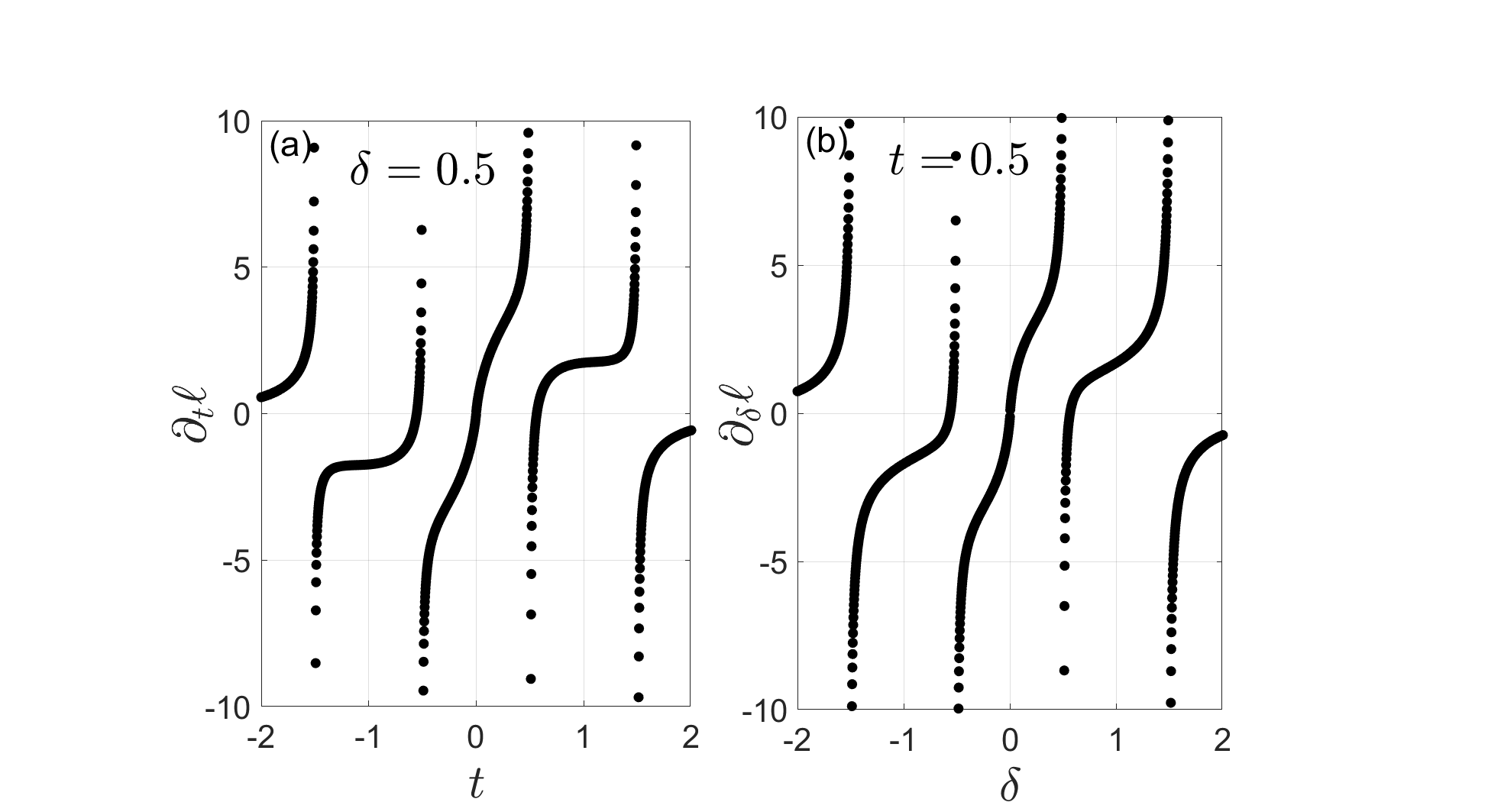}
	\caption{The derivatives of the length with respect to $t$ and $\delta$ in the complex energy plane. It can be seen that the derivatives are divergent at the topological phase transition.}
	\label{fig3}
\end{figure}

In Fig. (\ref{fig3}) (a) and (b), we plot numerically the derivations of the length of the curve with respect to $t$ and $\delta$ respectively.
It can be seen that the derivations of the length with respect to $t$ and $\delta$ are discontinuous and divergent in the boundary of the topological phase transition. The divergent points in the parameter space are consistent with the analytic result, $t=\pm\delta\pm 1$.

\subsection{Non-Hermitian Chern insulator model}

The 2D non-Hermitian Chern insulator model contains rich topological phases characterized by Chern number and vorticity.\cite{Kohei2} We first review briefly the basic results of this model and then compare them with those by the geometric criterion method. The Bloch Hamiltonian of the non-Hermitian Chern insulator is given by \cite{Ghatak}
\begin{equation}\label{CS1}
H(\mathbf{k})=\mathbf{h}(\mathbf{k})\cdot \mathbf{\sigma}
\end{equation}
where
\begin{eqnarray}\label{hhh1}
h_{x}(\mathbf{k}) &=& t\sin k_{x},\\
h_{y}(\mathbf{k}) &=& t\sin k_{y}-i\delta \\
h_{z}(\mathbf{k}) &=& m+t\cos k_{x}+t\cos k_{y},
\end{eqnarray}
where $t,m,\gamma\in \mathcal{M}^{3}_{\lambda}$ are real parameters and $t\in \mathbb{R}^+$. $(k_x,k_y)\in [-\pi,\pi]^2$. In the case of $\delta=0$, the model reduces to the well-known Hermitian Chern insulator,  which is characterized by the Chern number.\cite{Kane,Zhang,Chiu} The energy bands of the non-Hermitian Chern insulator are given by
\begin{equation}\label{EECS}
E_{\pm}=\pm\sqrt{m^2-\gamma^2+2t^2(1+\cos k_{x}\cos k_{y})+2mt(\cos k_{x}+\cos k_{y})-2it\delta\sin k_{y}}.
\end{equation}
The real and imaginary parts of the energy bands are obtained by
\begin{eqnarray}
E_{R} &=& \sqrt{\frac{\epsilon+\sqrt{\epsilon^2+\omega^2}}{2}}, \label{RIEB1a}\\
E_{I} &=& \sqrt{\frac{-\epsilon+\sqrt{\epsilon^2+\omega^2}}{2}}, \label{RIEB1b}
\end{eqnarray}
where
\begin{eqnarray}
\epsilon &=& 2t^2+m^2-\delta^2+2t^{2}\cos k_x\cos k_y +2mt(\cos k_x+\cos k_y),  \label{RIEB2a}\\
\omega &=& -2t\delta \sin k_y.  \label{RIEB2b}
\end{eqnarray}
The complex energy bands can be rewritten as
\begin{equation}\label{CEB2}
E_{\pm}=\pm (E_R+\sigma i E_I),
\end{equation}
where $\sigma=1$ for $\omega\geq 0$ and $\sigma=-1$ for $\omega < 0$.

It has been found that there exists two types of physical mechanisms associated with topological phases, which are gapped and gapless phases. The gapped phase is characterized by the Chern number defined by\cite{Kohei2}
The Chern number is defined by
\begin{equation}\label{BC1}
C_{\pm} = \frac{1}{2\pi}\int_{\textrm{BZ}}\Omega^{\pm}(\mathbf{k})d^{2}\mathbf{k}
\end{equation}
where $\Omega^{\pm}(\mathbf{k})$ is the Berry curvature, which is given by \cite{Annan}
\begin{equation}\label{BC2}
\Omega^{\pm}(\mathbf{k}) = \mp\frac{1}{2}\mathbf{\widehat{h}}\cdot
\left(\frac{\partial \mathbf{\widehat{h}}}{\partial k_x}\times\frac{\partial \mathbf{\widehat{h}}}{\partial k_y}\right)
\end{equation}
where $\mathbf{\widehat{h}}$ is the unit vector of $\mathbf{h}$.
The gapless phase is characterized by the vorticity, which is defined by \cite{Kohei2,Ghatak}
\begin{equation}\label{VT1}
\nu(\mathbf{k}_{EP})=\frac{1}{2\pi i}\oint_{\mathcal{C}(\mathbf{k}_{EP})}\nabla_{\mathbf{k}}\arg (E_{n}(\mathbf{k})-E_{m}(\mathbf{k})) \cdot d\mathbf{k}
\end{equation}
$\mathcal{C} (\mathbf{k}_{EP})$ in (\ref{VT1}) denotes a loop encircled the exceptional point in ZB.

The boundaries of the topological phases occur at the exceptional points in BZ, which corresponds to the closure of the complex energy band, $\left(E_{\pm}(\mathbf{k})\right)_{EP}$, where $(k_{x})_{EP}\in BZ^2$. Note that the energy bands in terms of (\ref{RIEB2a}) and (\ref{RIEB2b}), the exceptional points are located at $(k_{x})_{EP}=0,\pm\pi$, which leads to the boundaries of the topological phase transition in the parameter space,\cite{Kohei2}

\begin{eqnarray}\label{BC1}
|m|\leq |\gamma| \leq |m+2t| \quad \textrm{for} \quad m\geq -t \\
|m+2t|\leq |\gamma| \leq |m| \quad \textrm{for} \quad m\leq -t,
\end{eqnarray}
for $(k_{x})_{EP}=0$ and
\begin{eqnarray}
|m-2t| \leq |\gamma| \leq  |m|\quad \textrm{for} \quad m\geq t \\
|m|\leq |\gamma| \leq |m-2t| \quad \textrm{for} \quad m\leq t, \label{BC2}
\end{eqnarray}
for $(k_{x})_{EP}=\pm\pi$. The Chern number and the vorticity in (\ref{BC1}) characterize different topological phases, gapped and gapless phases.\cite{Kohei2}

Here we reexamine the topological phase diagram by the geometric criterion, in which the analytic behaviors of the lengths of the boundary of the bulk band in the complex energy plane play a crucial role in the topological phase transition. The curve of the boundary of the states in the complex energy plane for given parameters is defined by
\begin{equation}\label{PBS1}
\ell (\mathbf{\lambda}):=\int_{\partial \varepsilon^c}d\ell_{\varepsilon^c}=\int_{\partial \varepsilon^c}\left(\frac{\partial \ell}{\partial E_R}dE_R +\frac{\partial \ell}{\partial E_I}dE_I\right)
\end{equation}
where $\ell (\mathbf{\lambda}):=\partial \varepsilon^c$. However, there is no analytic function $\ell(E_R,E_I)$ to obtain the length of the boundary in the complex energy plane because the length in (\ref{PBS1}) is not well-defined in the Riemann integral for the 2D BZ, but it is well-defined in the Lebesgue integral. Thus, we can calculate the length by the numerical method called the ball pivoting algorithm.\cite{Fausto}

\begin{figure}[htbp]
	\centering
	\includegraphics[width=1.2\linewidth]{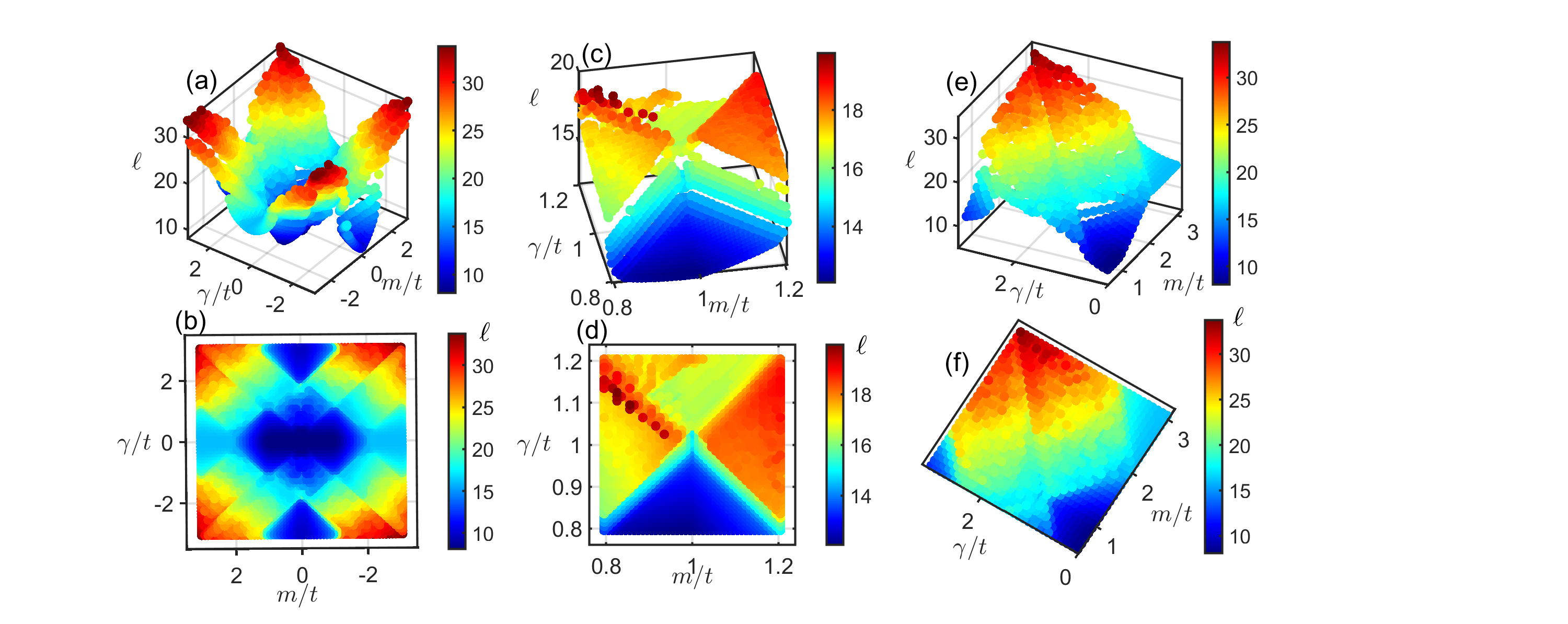}
	\caption{(a): The length of the boundary of the bulk band in the complex plane is plotted in the parameter space. (b): The surface of the length is projected to the parameter space, which exhibits the topological phase diagram and is exactly same to the analytic results from (\ref{BC1}) to (\ref{BC2}). (c): The local detail of the length near the boundary of the topological phase transition between the gapped to gapless phases, in which we can see the discontinuous length in the boundary of the phase transition. (d): The local detail of the length in (c) is projected on the parameter space. (e): The local detail of the length near the boundary of the topological phase transition between the gapless to gapless phase, in which we can see that the length in the boundary of the phase transition is continuous, but the derivative of the length with respect to the parameters is discontinuous. (f): The local detail of the length in (e) is projected on the parameter space. }
	\label{fig4}
\end{figure}

In Fig. \ref{fig4} (a) and (b), the length of the boundary of the bulk band in the complex energy plane are plotted in the parameter space.
We find that nonanalytic behaviors of the length of the boundary of the bulk band in the complex energy plane occur in the boundary of the topological phase transition in Fig. \ref{fig4} (a). When the length is projected to the parameter space we can see the topological phase diagram in the parameter space in Fig. \ref{fig4} (b), which is exactly same to the analytic results from (\ref{BC1}) to (\ref{BC2}). In Fig. \ref{fig4} (c), we plot the local detail of the length near the boundary of the topological phase transition between the gapped to gapless phases in the parameter space, in which we can see the discontinuous length in the boundary of the phase transition. The local detail of the length in (c) is projected on the parameter space shown in Fig. \ref{fig4} (d). Similarly, Fig. \ref{fig4} (e) shows the local detail of the length near the boundary of the topological phase transition between the gapless to gapless phases in the parameter space, in which we can see that the length in the boundary of the phase transition is continuous, but the derivative of the length with respect to the parameters is discontinuous. The local detail of the length in Fig. \ref{fig4} (e) is projected on the parameter space in Fig. \ref{fig4} (f).

From the geometric point of view, there is one region of the bulk band in the complex energy plane for the gapless phase but there are two regions for the gapped phase. The phase transition from the gapped to gapless phases implies the closure of energy-band gap, which corresponds to the two regions combining to one region in the complex energy plane. Thus the length of the boundary jumps when the phase transition from the gapped to gapless phases.

\section{Remarks and discussions}
The geometric criterion of the topological phase transition inspires us two fundamental issues.
What physical mechanism is behind of this geometric criterion and what physical observable is related to the length of the bulk band in the complex energy plane? These are worth studying further.
However, we can understand this geometric criterion from two points. One is the local-global correspondence in the complex energy plane, the other is the topological property of transformation between the BZ and the complex energy plane.

For the 2D BZ, the geometric method can be viewed as a generalized boundary-bulk correspondence in the complex energy plane. The boundary of the bulk band in the complex energy plane can be regarded as the complex edge states in the complex energy plane. The geometric criterion implies some local information from the edge states dominating the global topological properties of non-Hermitian systems. The discontinuous behaviors of the length come from two bands contact or separating in the complex energy plane, namely the complex energy band gap closes or opens.

We study the map of quantum states from the BZ to the complex energy plane. For given parameters, the map from the $BZ^{2}\times \mathcal{M}^{2}_{\lambda}$ to $\mathcal{E}^{c}$ can be represented as a transformation from the BZ to the complex energy plane,
\begin{equation}\label{TM1}
\left(
\begin{array}{c}
dE_{R} \\
dE_{I}
\end{array}
\right)=
J(\mathbf{k},\lambda)
\left(
\begin{array}{c}
dk_x \\
dk_y
\end{array}
\right),
\end{equation}
where $J(\mathbf{k},\lambda)$ is the transformation matrix, which is given by
\begin{equation}\label{JD1}
J(\mathbf{k},\lambda)=\left(
\begin{array}{cc}
\frac{\partial E_{R}}{\partial k_{x}} & \frac{\partial E_{R}}{\partial k_{y}} \\
\frac{\partial E_{I}}{\partial k_{x}} & \frac{\partial E_{I}}{\partial k_{y}}
\end{array} \right).
\end{equation}
We investigate the determinant of the transformation matrix $\det J(\mathbf{k},\lambda)$, namely the Jacobian determinant. When $\det J(\mathbf{k},\lambda)\neq 0$ for $\forall \mathbf{k} \in BZ$, the transformation is invertible. When there exists any $\mathbf{k} \in BZ$ such that $\det J(\mathbf{k},\lambda)= 0$, it implies that the transformation is not invertible. In other words, the transformation between the BZ and the complex energy plane is not one-to-one.

Similarly to the non-Hermitian SSH model, note that the relationship between the real and imaginary parts of the energy band in (\ref{RIEB1a}) and (\ref{RIEB1b}), we have
$E^{2}_{R}-E^{2}_{I}=\epsilon$ and $E^{2}_{R}E^{2}_{I}=\omega^{2}/4$ , namely
\begin{eqnarray}
E_{R}^{2}-E_{I}^2 &=& 2-\delta^{2}+m^{2}+2\cos k_{x}\cos k_{y}+2m(\cos k_{x}+\cos k_{y}), \label{ERI3} \nonumber \\
E_{R}E_{I} &=& \delta\sin{k_{y}}. \label{ERI4}
\end{eqnarray}
where we set $t=1$ for convenience without loss of generality. By using derivative with respect to $k_x$ and $k_y$ to (\ref{ERI3}) and (\ref{ERI4}), we have
\begin{eqnarray}\label{DERIx}
E_{R}\frac{\partial E_R}{\partial k_x}-E_{I}\frac{\partial E_I}{\partial k_x} &=& -\sin k_{x}\cos k_{y}-m\sin k_{x}, \\
E_{I}\frac{\partial E_R}{\partial k_x}+E_{R}\frac{\partial E_I}{\partial k_x}&=& 0.
\end{eqnarray}
and
\begin{eqnarray}\label{DERIy}
E_{R}\frac{\partial E_R}{\partial k_y}-E_{I}\frac{\partial E_I}{\partial k_y} &= & -\cos k_{x}\sin k_{y}-m\sin k_{y}, \\
E_{I}\frac{\partial E_R}{\partial k_y}+E_{R}\frac{\partial E_I}{\partial k_y} &= & \delta \cos k_{y}.
\end{eqnarray}
We can rewrite (\ref{DERIx}) and (\ref{DERIy}) to
\begin{equation}\label{DERIx3}
\left(
\begin{array}{cc}
E_{R} & -E_{I} \\
E_{I} & E_{R}
\end{array}
\right)
\left(
\begin{array}{c}
\frac{\partial E_R}{\partial k_x} \\
\frac{\partial E_I}{\partial k_x}
\end{array}
\right)=\left(
\begin{array}{c}
-(\cos k_{y}+m)\sin k_{x} \\
0
\end{array}
\right),
\end{equation}
and
\begin{equation}\label{DERIy3}
\left(
\begin{array}{cc}
E_{R} & -E_{I} \\
E_{I} & E_{R}
\end{array}
\right)
\left(
\begin{array}{c}
\frac{\partial E_R}{\partial k_y} \\
\frac{\partial E_I}{\partial k_y}
\end{array}
\right)=\left(
\begin{array}{c}
-(\cos k_{x}+m)\sin k_{y} \\
\delta\cos k_y
\end{array}
\right).
\end{equation}
For the gapped regions, $E=\sqrt{E_{R}^{2}+E_{I}^{2}}\neq 0,\forall \mathbf{k}\in BZ $, there exists the inverse of the energy matrix in (\ref{DERIx3}) and(\ref{DERIy3}). Thus, the derivatives of the real and imaginary energy bands can be expressed as
\begin{equation}\label{DERIx4}
\left(
\begin{array}{c}
\frac{\partial E_R}{\partial k_x} \\
\frac{\partial E_I}{\partial k_x}
\end{array}
\right)=
\left(
\begin{array}{cc}
E_{R} & -E_{I} \\
E_{I} & E_{R}
\end{array}
\right)^{-1}
\left(
\begin{array}{c}
-(\cos k_{y}+m)\sin k_{x} \\
0
\end{array}
\right),
\end{equation}
and
\begin{equation}\label{DERIy4}
\left(
\begin{array}{c}
\frac{\partial E_R}{\partial k_y} \\
\frac{\partial E_I}{\partial k_y}
\end{array}
\right)=
\left(
\begin{array}{cc}
E_{R} & -E_{I} \\
E_{I} & E_{R}
\end{array}
\right)^{-1}
\left(
\begin{array}{c}
-(\cos k_{x}+m)\sin k_{y} \\
\delta\cos k_y
\end{array}
\right),
\end{equation}
where the inverse of the matrix is given by
\begin{equation}\label{DERIy2}
\left(
\begin{array}{cc}
E_{R} & -E_{I} \\
E_{I} & E_{R}
\end{array}
\right)^{-1}=\frac{1}{E^{2}}
\left(
\begin{array}{cc}
E_{R} & E_{I} \\
-E_{I} & E_{R}
\end{array}
\right).
\end{equation}
Consequently, the derivatives of the real and imaginary parts of the energy bands with respect to $k_x$ and $k_y$ are given by
\begin{equation}\label{DERIx6}
\left(
\begin{array}{c}
\frac{\partial E_R}{\partial k_x} \\
\frac{\partial E_I}{\partial k_x}
\end{array}
\right)=\frac{1}{E^{2}}
\left(
\begin{array}{c}
-E_{R}(\cos k_{y}+m)\sin k_{x} \\
E_{I}(\cos k_{y}+m)\sin k_{x}
\end{array}
\right),
\end{equation}
and
\begin{equation}\label{DERIy6}
\left(
\begin{array}{c}
\frac{\partial E_R}{\partial k_y} \\
\frac{\partial E_I}{\partial k_y}
\end{array}
\right)=\frac{1}{E^{2}}
\left(
\begin{array}{c}
\delta E_{I}\cos k_{y}-E_{R}(\cos k_x+m)\sin k_{y} \\
\delta E_{R}\cos k_{y}+E_{I}(\cos k_x+m)\sin k_{y}
\end{array}
\right).
\end{equation}
By substituting (\ref{DERIx6}) and (\ref{DERIy6}) into (\ref{JD1}), the Jacobian determinant of the transformation from the BZ to the complex energy plane can be obtained
\begin{equation}\label{JD3}
\det J(\mathbf{k},\lambda)=-\frac{\delta (m+\cos k_y)\sin k_x \cos k_y}{E^2}.
\end{equation}

It should be noted that when the Jacobian determinant is zero, $\det J(\mathbf{k},\lambda)=0$, it yields the solutions either $k_{y}= \pm \frac{\pi}{2}$ or $k_{x}=0,\pm \pi$ or $\cos k_{y}=-m$. We find that the quantum states of the zero-Jacobian determinant correspond to the edge states of the boundary of the bulk bands in the complex energy plane. That is why the boundary states dominate the topological invariant. \cite{Annan2}
Thus, the zero-Jacobian determinant gives a signal for the topological invariance for non-Hermitian systems.

The zero-Jacobian determinant means the transformation from the BZ to the complex energy plane is not one-to-one, namely the boundary states of the bulk bands in the complex energy plane are degenerate states in the BZ, which is consistent with the previous results on the topological order in the ground states.\cite{Wen}

In general, the length and its derivative respect with the parameters in the complex energy plane contains the information coming from non-Hermiticity of systems. The length contains the real and imaginary parts of the energy band. The imaginary part plays some roles in some dissipative effects of non-Hermitian systems.
The dimension of the length of the bulk band in the complex energy plane is energy. This reveals the geometric variables in energy space.
From the mathematical point of views, the geometric variables are robust and independent of the coordinate systems. This is why the length in the complex energy plane can detect the global topological invariants in non-Hermitian systems.
The analytic behavior of the length in the complex energy plane measures the geometric properties in the energy space for non-Hermitian systems.

It should be remarked that the length of the bulk band in the complex energy plane for the 2D BZ is calculated numerically by the ball pivoting algorithm,\cite{Fausto,Annan2} by which we can yield the boundary of the bulk band in the complex energy plane.
The basic steps of the ball pivoting algorithm are that (1) for given a 2D region, inputting a ball (actually circle for 2D regions) with the radius $r$ as an initial value, (2) rotating the circle along the boundary of the region, (3) counting the number of the rotating circle, and (4) calculating the length of the boundary of the region. The precision of the algorithm depends on the radius of the circle. Reducing the numerical errors pay the computer time. However, for the boundary of the topological phase transition, this numerical algorithm is efficient and stable because the topological phase is robust for the parameter deformation. The numerical errors do not change the results because the topological phase transition depends only on the analytic behaviors of the length.\cite{Annan2}
In practice, we can tune the radius of the circle to increase the precision of the algorithm by paying the computer time.
Moreover, this geometric criterion of the topological phase transition can be applied to more generic non-Hermitian systems because it does not involve the symmetry of systems provided the energy band we consider is that near Fermi energy or the Fermi surface.

\section{Conclusions and outlooks}
The topological phase plays an important role in quantum technology due to its robustness against the parameter deformation.\cite{Chiu,Zhang,Wen,Longhi} In particular, the non-Hermitian systems have rich topological phases, which inspire many fundamental issues, such as what physical observable behind the topological phase and how to tune efficiently the topological phase transition for non-Hermitian systems.\cite{Gong,Kohei,Annan}

We propose a geometric criterion to detect the topological phase transition for non-Hermitian systems.
We define a map by the non-Hermitian Hamiltonian from the BZ to the complex energy plane for given parameters.
The bulk energy band forms a curve in the complex energy plane for the 1D BZ.
We find that the derivatives of the length with respect to the parameters in the complex energy plane discontinuous in the boundary of the topological phase transition, which provides a signal for the topological phase transitions. We demonstrate the phase diagrams of the non-Hermitian SSH model based on this geometric criterion.
For the 2D BZ, the bulk band forms a 2D region in the complex energy plane.
The length of the boundary of the bulk band in the complex energy plane plays a crucial role to detect the topological phase transition. The length
is discontinuous in the topological phase boundaries between the gapped and gapless phases, whereas the derivative of the length with respect to parameters is discontinuous in
the topological phase transitions between the gapless and gapless. We demonstrate this geometric criterion
by repeating numerically the phase diagrams of the non-Hermitian SSH and Chern insulator models.

This geometric criterion can be regarded as a generalized boundary-bulk correspondence in the complex energy plane, which reveals some states in the complex energy plane dominating the global topological invariants for non-Hermitian systems. The length of the bulk band in the complex plane is a geometric variable with the energy dimension.
This tells us that the geometric length associated with the complex energy scale dominates the topological invariant for non-Hermitian systems. This implies that the topological invariants still occur even though there exists the non-Hermiticity for non-Hermitian systems. The non-Hermiticity is associated with the non-equilibrium or dissipative phenomena in non-Hermitian systems.

In principle, this geometric method can be generalized to study more generic non-Hermitian systems because this method does not depend on symmetry of systems.
As long as we take the energy band near Fermi energy into account to investigate the analytic properties of the length in the complex energy plane,
this geometric method should be available because the energy band near Fermi energy plays a crucial role in physical properties of systems.
Actually, we found that the topological invariant can be characterized by the pattern invariant of the boundary states in the complex energy plane mapped to the BZ. \cite{Annan2}. We also studied the topological phase transition of the Aubry-Andre-Harper (AAH) model based on this geometric criterion.\cite{Annan3} The Hamiltonian of the AAH model is a $3\times3$ matrix with the 1D BZ. We found that the derivative of the total length of the three energy bands with respect to the parameter is also discontinuous at the topological phase transition, which corresponds to the curves of the three bands in the complex energy plane changing from three separate curves to a connected curve with the variation of the parameter, and the Jacobian determinant of the transformation between the BZ and the complex energy plane being zero. These results imply that the geometric criterion is still available for more generic models. \cite{Annan3}
This geometric method reveals a connection between the geometric variable in the complex energy plane and the topological phases.

Physically, the length in the complex energy plane is associated with the density of states of the energy band and its degeneracy.
In general, the degenerate states imply that the Bloch wave function is localized, which is related to the electron and heat transportation.\cite{Mermin} Interestingly, due to the non-Hermiticity, the length in the complex energy plane contains some dissipative effects.
In principle, the dissipative or open system contains system and its environment, which is described by the density matrix approach and Liouville-von Neumann equation,\cite{Breuer} especially for quantum optical systems. However, in condensed matter physics, the environment effects are usually effectively included in the non-Hermitian terms of the Hamiltonian by parametrization, in which the diagonal terms describe the energy loss or gain of the system and the off-diagonal terms describe the non-equivalent hoppings between two nearest sides of the lattice model.\cite{Moiseyev}
This non-Hermiticity of the Hamiltonian yields complex energy bands. In other words, the length of the curve in the complex energy plane
contains the real and imaginary contributions of the complex energy bands. The imaginary part reflects the dissipative effect of the non-Hermitian systems.
The slope of the curve in the complex energy plane is the ratio of the imaginary and real parts of the complex energy band.
What detailed features of the length, such as  real or imaginary part, or both, dominating the phase transition are still unknown, which could also depend on the concrete model. This is an interesting issue for non-Hermitian systems. We found that the Chern number generalized to the complex Chern number corresponds to the quantum Hall conductance generalized to quantum Hall admittance for the non-Hermitian Dirac model.\cite{Annan}

The robust property of the topological phase hides some unknown physical phenomena in non-Hermitian systems.\cite{Annan2}.
How to capture the dissipative effects from the variation of the length in the complex energy plane is worth studying further.

\begin{acknowledgments}
The authors are grateful to a referee for valuable comments and suggestions.
\end{acknowledgments}


\bibliography{apssamp}

\end{document}